\documentclass[aps,prc,reprint,superscriptaddress,preprintnumbers,showpacs,nofootinbib,a4paper]{revtex4-1}

\usepackage{verbatim}
\usepackage[latin1]{inputenc}
\usepackage{graphicx,amssymb} 

\usepackage{dcolumn}
\usepackage{bm}
\usepackage{color}%
\usepackage[bookmarks=false, pdfborder={0 0 0}, colorlinks=true, linkcolor=blue, citecolor=blue, filecolor=blue, urlcolor=blue]{hyperref}
\usepackage[all]{hypcap}

\graphicspath{{./Figures-EPS/}{../Figures-EPS/}{.}{../LiteratureReview/}}

\newcommand{\hzdr}{\affiliation{Helmholtz-Zentrum Dresden-Rossendorf (HZDR), Bautzner Landstr. 400, D-01328 Dresden, Germany}}
\newcommand{\tu}{\affiliation{Technische Universit\"at Dresden, Institute of Nuclear and Particle Physics, D-01062 Dresden, Germany}}

\newcommand{\pd}{\affiliation{Istituto Nazionale di Fisica Nucleare (INFN), Sezione di Padova, I-35131 Padova, Italy}}

\begin{document}

\title{Strength of the $\bm{E_{\text{p}}=1.842}$\,MeV resonance in the $\bm{^{40}}$Ca(p,$\bm{\gamma}$)$\bm{^{41}}$Sc reaction revisited} 

\thispagestyle{empty}

\author{Konrad Schmidt}\hzdr\tu 			
\author{Shavkat Akhmadaliev}\hzdr 			
\author{Michael Anders}\hzdr\tu 			
\author{Daniel Bemmerer}\email{d.bemmerer@hzdr.de}\hzdr	
\author{Antonio Caciolli}\pd 				
\author{Mirco Dietz}\hzdr\tu				
\author{Zolt\'an Elekes}\hzdr				
\author{Arnd R. Junghans}\hzdr 				
\author{Marie-Luise Menzel}\hzdr\tu 			
\author{Ronald Schwengner}\hzdr 			
\author{Andreas Wagner}\hzdr 				
\author{Kai Zuber}\tu

\begin{abstract}

The strength of the $E_{\rm p} = 1.842$\,MeV resonance in the $^{40}$Ca(p,$\gamma$)$^{41}$Sc reaction is determined with two different methods: First, by an absolute strength measurement using calcium hydroxide targets, and second, relative to the well-determined strength of the resonance triplet at $E_\alpha$ = 4.5\,MeV in the $^{40}$Ca($\alpha$,$\gamma$)$^{44}$Ti reaction. The present new value of $\omega\gamma=(0.192\pm0.017)$\,eV is 37\% (equivalent to $3.5\sigma$) higher than the evaluated literature value. In addition, the ratio of the strengths of the 1.842\,MeV $^{40}$Ca(p,$\gamma$)$^{41}$Sc and 4.5\,MeV $^{40}$Ca($\alpha$,$\gamma$)$^{44}$Ti resonances has been determined to be $0.0229\pm0.0018$. The newly corrected strength of the 1.842-MeV resonance can be used in the future as a normalization point for experiments with calcium targets.

\end{abstract}

\maketitle

\section{Introduction}
\label{sec:Introduction}

Precise values for selected resonance strengths may serve as normalization points for nuclear reaction experiments~\cite{Iliadis2007}. This is particularly true for nuclei of astrophysical interest, where in several cases precision cross section data are needed in order to constrain astrophysical scenarios. One example are mirror nuclei such as $^{40}$Ca that are included in the $\alpha$-rich freezeout process believed to be responsible for the production of the supernova nuclide $^{44}$Ti~\cite{Diehl2010}.

The $E_{\rm p}=1.842$\,MeV resonance in the $^{40}$Ca(p,$\gamma$)$^{41}$Sc reaction provides a useful normalization point for experiments addressing the $\alpha$-rich freezeout, because it is relatively strong, easily accessible by a proton beam and generally in the astrophysically relevant energy range. This resonance populates the $E_{\rm x}=2882$\,keV, 7/2$^+$, seventh excited state in $^{41}$Sc, which, in turn, decays with $>$99.9\% probability by $\gamma$-ray emission to the ground state~\cite{Cameron2001}. $^{41}$Sc is $\beta^+$ unstable with a half life of 0.6\,s and a positron endpoint energy of 5.473\,MeV for the strongest decay branch.

The strength of the $E_{\rm p} = 1.842$\,MeV resonance has been measured several times in the past (Table~\ref{table:LiteratureReview}). In the framework of networks of (p,$\gamma$) resonance strengths involving several different nuclei, its value was first determined by Butler~\cite{Butler1961} on calcium oxide targets using in-beam $\gamma$-ray spectroscopy with NaI detectors and detecting the positrons from the decay of $^{41}$Sc. Youngblood {\it et al.}~\cite{Youngblood1968} devoted considerable effort to obtain a pure metallic calcium target and measured the resonance strength in an absolute way again by positron counting. A third absolute measurement was performed by Kozub {\it et al.}~\cite{Kozub1977}, again on metallic calcium targets but using in-beam $\gamma$ spectrometry with germanium detectors. Finally, as an ancillary result of an experiment aiming to study the $^{40}$Ca($\alpha$,$\gamma$)$^{44}$Ti reaction, Robertson {\it et al.} report an absolute value of the $^{40}$Ca(p,$\gamma$)$^{41}$Sc resonance strength~\cite{Robertson2012}. Relative resonance strengths measurements have been reported by Engelbertink and Endt~\cite{Engelbertink1966} and by Paine and Sargood~\cite{Paine1979}.

Here, a new measurement of the resonance strength is presented. To this end, data taken in the framework of a recent $^{40}$Ca($\alpha$,$\gamma$)$^{44}$Ti experiment~\cite{Schmidt2013} are re-analyzed with a view to extract the strength of the $E_{\rm p} = 1.842$\,MeV resonance in $^{40}$Ca(p,$\gamma$)$^{41}$Sc. The sought for resonance strength is determined both absolutely and relative to the recently redetermined ($\alpha$,$\gamma$) strength. 

\begin{table*}
  \caption{Strength of the $E_{\rm p} = 1.842$\,MeV resonance, from the literature and from this work.\label{table:LiteratureReview}}
  \begin{ruledtabular}
  \begin{tabular}{D{.}{.}{3}@{$\,\pm\,$}D{.}{.}{3}lll}
    \multicolumn{2}{c}{$\omega\gamma$ (eV)}	& Reference	& Target	& Technique \\
    \hline\noalign{\smallskip}
    0.15	& \multicolumn{1}{l}{$\!^{0.15}_{0.08}$}
			& Butler~\cite{Butler1961}
				& CaO
					& absolute; in-beam $\gamma$ spectrometry and $^{41}$Sc $\beta^+$-counting \\
    0.13	& 0.02	& Engelbertink and Endt~\cite{Engelbertink1966}
				& Ca$_3$(PO$_4$)$_2$, CaSO$_4$
					& relative to $^{31}$P(p,$\gamma$)$^{32}$S, $^{32}$S(p,$\gamma$)$^{33}$Cl resonances \\
    0.193	& 0.047 & Youngblood {\it et al.}~\cite{Youngblood1968}
				& metallic Ca
					& absolute; $^{41}$Sc $\beta^+$-counting \\
    0.14	& 0.02	& Kozub {\it et al.}~\cite{Kozub1977}
				& metallic $^{40}$Ca
					& absolute; in-beam $\gamma$ spectrometry \\
    0.140	& 0.025	& Paine and Sargood~\cite{Paine1979}
				& CaO on Al
					& relative to $^{27}$Al(p,$\gamma$)$^{28}$Si resonance \\
    0.14	& 0.02	& Robertson {\it et al.}~\cite{Robertson2012}
				& metallic Ca
					& absolute; in-beam $\gamma$ spectrometry \\
    0.192 	& 0.017	& present work
				& Ca(OH)$_2$
					&  both absolute and relative to $^{40}$Ca($\alpha$,$\gamma$)$^{44}$Ti \\
  \end{tabular}
  \end{ruledtabular}
\end{table*}

\section{Experiment}
\label{sec:Experimental}

For the present purposes, the data from the scans of the $E_{\rm p} = 1.842$\,MeV resonance for two different targets called \#31 and \#32 used for a study of the $^{40}$Ca($\alpha$,$\gamma$)$^{44}$Ti reaction~\cite{Schmidt2013} are re-analyzed. The experiment has been performed at the 3\,MV Tandetron accelerator of Helmholtz-Zentrum Dresden-Rossendorf (HZDR). 

Targets consisting of calcium hydroxyde with natural isotopic composition on a tantalum backing were irradiated at an angle of 55$^\circ$ tilted to the beam. The $\gamma$ rays from the reaction under study were detected by two escape-suppressed high-purity germanium (HPGe) detectors placed at angles of 55$^\circ$ and 90$^\circ$ to the beam direction, respectively. Further details on the experimental setup have been reported previously~\cite{Schmidt2013}.

\subsection{Analysis method\label{sub:Analysis method}}

The resonance strength $\omega\gamma$ is related to the proton, photon, and total widths $\Gamma_p$, $\Gamma_\gamma$, and $\Gamma$ of the resonance under study by the following equation:

\begin{eqnarray}
  \omega\gamma	& = & \frac{2J+1}{(2j_1+1)(2j_2+1))} \frac{\Gamma_p\Gamma_\gamma}{\Gamma} \\	\label{eq:omegagamma}
		& = & \frac{1}{(2j_1+1)(2j_2+1))} S(p,\gamma)					\label{eq:Spgamma}
\end{eqnarray}
The statistical factor $\omega$ depends on the total angular momenta $j_1$, $j_2$, and $J$ of projectile, target, and resonance. In earlier works, commonly an alternative expression is used, i.e. $S(p,\gamma)$ = $(2J+1)\Gamma_p\Gamma_\gamma/\Gamma$. For the reaction studied here, $S(p,\gamma)$ = 2$\omega\gamma$. 
For a target of infinite thickness, the experimental yield $Y_{\infty}$ as a function of $\omega\gamma$ is then given by the following relation~\cite{Iliadis2007}:
\begin{equation}\label{eq:Yield_omega_gamma}
  Y_{\infty} = \frac{\lambda_{\rm res}^2}{2} \frac{\omega\gamma}{\varepsilon_{\rm eff}^{\rm H}(E_{\rm res})}
\end{equation}
where $\lambda_{\rm res}$ is the de Broglie wavelength at the resonance energy $E_{\rm res}$ and $\varepsilon_{\rm eff}^{\rm H}$ is the effective stopping power for the hydrogen beam. 

The yield $Y_{\infty}$ critically depends on the stoichiometric composition of the target. Assuming the target to be of the stoichiometry CaO$_x$H$_y$, the stoichiometric parameters $x$ and $y$ affect the effective stopping power $\varepsilon_{\rm eff}^{\rm i}$ with $i$ $\in$ \{H, He\} in the following way:

\begin{equation}\label{eq:EffectiveStoppingPower}
  \varepsilon_{\rm eff}^{\rm i}(E) = \frac{f_B^{\rm i}(E)}{\eta_{40}} \cdot \left( \varepsilon_{\rm Ca}^{\rm i}(E) + x \cdot \varepsilon_{\rm O}^{\rm i}(E) + y \cdot \varepsilon_{\rm H}^{\rm i}(E) \, \right)
\end{equation}
In this relation, 
$\varepsilon_{\rm Ca}^{\rm i}(E)$ is the stopping power of ion $i$ in calcium,
$\varepsilon_{\rm O}^{\rm i}(E)$ in solid oxygen, and
$\varepsilon_{\rm H}^{\rm i}(E)$ in solid hydrogen.
The isotopic ratio of $^{40}$Ca in natural calcium is assumed to be $\eta_{40} = (96.94\pm0.03)$\%~\cite{Coplen2002}. The correction factor $f_B^{\rm H}(1.8$\,MeV$)=0.983$ takes into account the slight deviations from Bragg's stopping power summation rule. $f_B^{\rm H}$ has been estimated using the so-called core and bond approach~\cite{Ziegler1988} for stoichiometric Ca(OH)$_2$. For helium ions at a laboratory energy of 4.5\,MeV, $f_B^{\rm He}(4.5$\,MeV$)=0.997$~\cite{Schmidt2013}.

If one limits the experiment to just one target material, there are in principle two possible approaches to determine the resonance strength:
\begin{enumerate}
\item The experimental yield $Y_{\infty}$ is measured, and the stoichiometric parameters $x$ and $y$ in Eq.~(\ref{eq:EffectiveStoppingPower}) are determined in an absolute manner. The resonance strength then directly follows from Eq.~(\ref{eq:Yield_omega_gamma}).
\item Two different resonances, e.g. a (p,$\gamma$) and an ($\alpha$,$\gamma$) resonance on the same target nucleus, are studied in the same target, determining the experimental yields for each of them separately. The ratio of resonance strengths is then determined as follows:
\begin{equation}\label{eq:Ratios_omega_gamma}
  \frac{\omega\gamma({\rm p},\gamma)}{\omega\gamma(\alpha,\gamma)} = 
  \frac{Y_{\infty}({\rm p},\gamma)}{Y_{\infty}(\alpha,\gamma)}
  \frac{\lambda_{\rm res}^2(\alpha,\gamma)}{\lambda_{\rm res}^2({\rm p},\gamma)}
  \frac{\varepsilon_{\rm eff}^{\rm H}(E_{\rm res})}{\varepsilon_{\rm eff}^{\rm He}(E_{\rm res})}.
\end{equation}
As the ratio of effective stopping powers $\varepsilon_{\rm eff}^{\rm H}(E_{\rm res})/\varepsilon_{\rm eff}^{\rm He}(E_{\rm res})$ is usually only weakly dependent on the stoichiometry, this relative approach obviates the need to determine the target stoichiometry. 
\end{enumerate}

In the present work, both these approaches are used. Alternative approaches include relative measurements using either two resonances on different target nuclei both included in the same chemical compound~\cite{Engelbertink1966} or two different chemical compounds deposited subsequently on the same target backing~\cite{Paine1979}. However, different from approach (2) above these approaches still retain the dependence on the knowledge of the stoichiometric composition of each of the two compounds used.  

\subsection{Yield determination}

\begin{figure}
 \includegraphics[width=1.0\linewidth]{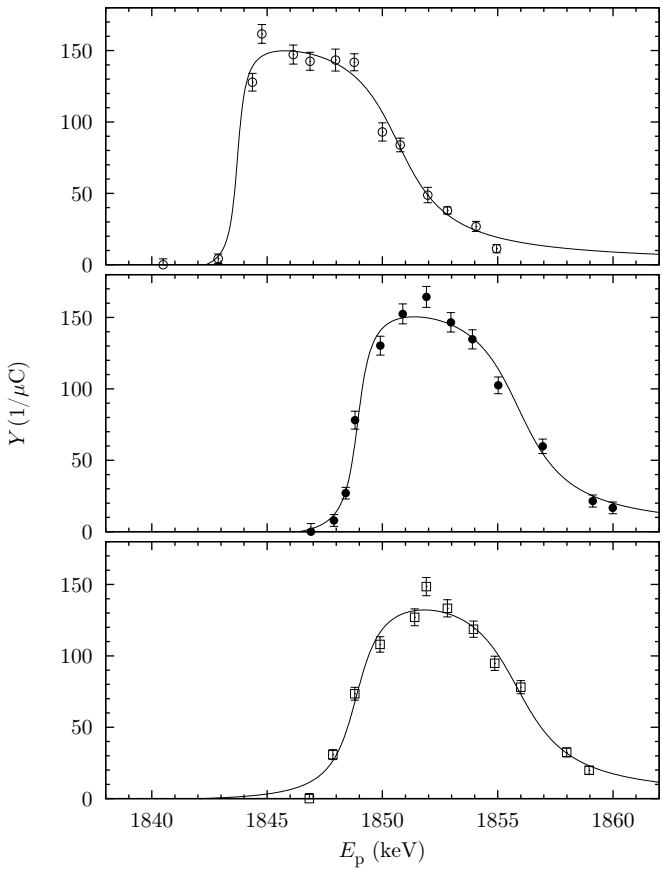}
 \caption{Experimental yield of the 2882\,keV $\gamma$ ray in target \#31 scans before first (open circles), between first and second (closed circles), and after second (open sqares) part of the $\alpha$-beam irradiation, as a function of proton beam energy. Lines are results of fits using Eq.~(\ref{eq:YieldCurveUnsymmmetric}). The proton energy shift in the two lower panels is due to an impurity layer buildup on the target. This layer also increases the beam energy straggling, which is seen by somewhat flatter slopes.}
 \label{fig:ProtonScanProfile}
\end{figure}

The resonance under study decays by $>$99.9\% by emission of a 2882\,keV $\gamma$ ray to the ground state in $^{41}$Sc~\cite{Cameron2001}. The experimental yield can thus be determined as a function of beam energy over the entire target width by observing this $\gamma$ ray. Two proton beam scans have been performed for target \#32: before and after the $\alpha$-beam irradiation~\cite{Schmidt2013}. For target \#31, the $\alpha$-beam irradiation was interrupted by an additional scan, so that there are three scans (Fig.~\ref{fig:ProtonScanProfile}).

The targets used here are rather narrow, with an energetic thickness of just 7.5\,keV for the proton beam. Therefore, the yield on the resonance plateau does not correspond to $Y_{\infty}({\rm p},\gamma)$ but must instead be extrapolated~\cite{Iliadis2007}. The yield as a function of proton energy $E_{\rm p}$ for a target of finite thickness~\cite{Fowler1948} is
\begin{eqnarray}\label{eq:YieldCurveFinite}
  Y(E_{\rm p}) = \frac{\lambda_{\rm res}^2}{2\pi}\,\frac{\omega\gamma}{\varepsilon_{\rm eff}^{\rm H}}\bigg(
		&	&	\arctan\frac{E_{\rm p}-E_{\rm res}}{\Gamma/2} \nonumber \\
		&	&	-\arctan\frac{E_{\rm p}-E_{\rm res}-\Delta E}{\Gamma/2}\bigg).
\end{eqnarray}
This yield curve can then be used to fit the measured yields in Fig.~\ref{fig:ProtonScanProfile}. However, due to beam energy straggling inside the target~\cite{Iliadis2007}, the slope of the right falling edge is less steep. In order to distinguish between the left and the right edge the two $\Gamma$'s in Eq.~(\ref{eq:YieldCurveFinite}) have been replaced with $\Gamma_{\rm left}$ and $\Gamma_{\rm right}$:
\begin{eqnarray}\label{eq:YieldCurveUnsymmmetric}
  Y(E_{\rm p}) = \frac{\lambda_{\rm res}^2}{2\pi}\,\frac{\omega\gamma}{\varepsilon_{\rm eff}^{\rm H}}\bigg(
		&	&	\arctan\frac{E_{\rm p}-E_{\rm res}}{\Gamma_{\rm left}/2} \nonumber \\
		&	&	-\arctan\frac{E_{\rm p}-E_{\rm res}-\Delta E}{\Gamma_{\rm right}/2}\bigg).
\end{eqnarray}
The measured yield curves are very well described by Eq.~(\ref{eq:YieldCurveUnsymmmetric}) (solid lines in Fig.~\ref{fig:ProtonScanProfile}), and the resonance strength $\omega\gamma$ is obtained directly from these fits (Table~\ref{table:ResonanceStrengths}). The statistical uncertainty is 6.2\% (6.3\%) for target \#31 (\#32). The combined value has a statistical uncertainty of 4.4\%.

\begin{table}
  \caption{Absolute resonance strengths $\omega\gamma({\rm p},\gamma)$ of targets \#31 and \#32. The first uncertainty given is purely statistical. For the average values of each target, the largest uncertainty of the three (two) proton scans (Fig.~\ref{fig:ProtonScanProfile}) of target \#31 (\#32) is adopted. For the combination of target \#31 and \#32, the uncertainty of the weighted mean (4.4\%) and the common systematic uncertainty (Table~\ref{table:UncertaintiesAbsolute}) are given.\label{table:ResonanceStrengths}}
  \begin{ruledtabular}
  \begin{tabular}{llD{.}{.}{1.3}@{$\,\pm\,$}l}
    \multicolumn{2}{c}{Target scan}			& \multicolumn{2}{c}{$\omega\gamma({\rm p},\gamma)$ (eV)} \\
    \hline\noalign{\smallskip}
    \#31	& before first $\alpha$-beam irradiation	& 0.192	& 0.012	\\
		& between $\alpha$-beam irradiations		& 0.199	& 0.008	\\
		& after second $\alpha$-beam irradiation	& 0.187	& 0.008	\\
    \multicolumn{2}{l}{Average \#31}				& 0.192	& 0.012	\\
    \hline\noalign{\smallskip}
    \#32	& before $\alpha$-irradiation			& 0.200	& 0.008	\\
		& after $\alpha$-irradiation			& 0.184	& 0.012	\\
    \multicolumn{2}{l}{Average \#32}				& 0.192	& 0.012	\\
    \hline\noalign{\smallskip}
    \multicolumn{2}{l}{\#31 and \#32 combined}			& 0.192	& 0.008 $\pm$ 0.015	\\
  \end{tabular}
  \end{ruledtabular}
\end{table}

\section{Results}
\label{sec:Results}

\subsection{Absolute determination of the strength of the 1.842-MeV resonance in $\bm{^{40}}$Ca(p,$\bm{\gamma}$)$\bm{^{41}}$Sc}
\label{sub:AbsoluteStrength}

For the absolute determination of the resonance strength, the stoichiometry of the targets has to be known. It has been determined previously for the two samples under study here~\cite{Schmidt2013} in two different ways: First, with an elastic recoil detection (ERD) analysis for a sample target from the same production batch, and second, by the analysis of the primary $\gamma$ rays from the $^{16}$O(p,$\gamma$)$^{17}$F reaction. Both methods gave consistent results~\cite{Schmidt2013}, and finally a stoichiometry of Ca(OH)$_{1.88\pm0.21}$ is obtained, which is consistent with calcium hydroxyde~\cite{Schmidt2013}. The stoichiometry contributes 5.9\% to the uncertainty of the resonance strength, half of the total error budget (Table~\ref{table:UncertaintiesAbsolute}).

The $\gamma$-ray angular distribution of the 2882\,keV $\gamma$ rays detected here has been measured previously by three independent groups~\cite{Youngblood1968,Rabin1973,Kozub1977}. Using the coefficients of Kozub {\it et al.}~\cite{Kozub1977}, which agree with Youngblood {\it et al.}~\cite{Youngblood1968} and Rabin~\cite{Rabin1973}, the ratio of angular distribution corrections at 90$^\circ$ and 55$^\circ$ results in $Y(90^\circ)/Y(55^\circ)|_{\rm literature}=0.703\pm0.027$. The present experimental ratio of yields for the 90$^\circ$ and the 55$^\circ$ detectors is $Y(90^\circ)/Y(55^\circ)|_{\rm present\,work}=0.674\pm0.019$, which confirms the correctness of the literature angular distribution. For the determination of $\omega\gamma({\rm p},\gamma)$, the present yields are corrected with the literature~\cite{Kozub1977} angular distribution, adding 3.8\% uncertainty on $\omega\gamma$.

The $\gamma$-ray detection efficiency has already been determined previously using calibrated radioactive sources and relative yields from the $^{27}$Al(p,$\gamma$)$^{28}$Si reaction, with an uncertainty of 2.3\% at 2882\,keV~\cite{Schmidt2013}.
The normalization of the stopping power~\cite{Ziegler2010} contributes another 1.4\% uncertainty.
The beam current was measured with a calibrated current integrator, and secondary electrons from the target were suppressed using a negatively charged tube just in front of the target, giving 1\% uncertainty for the beam intensity~\cite{Schmidt2013}.

Finally, the absolute resonance strength determined here is $\omega\gamma({\rm p},\gamma)=(0.192\pm0.017)$\,eV, with the error resulting from a quadratic combination of systematic and statistical uncertainties.

\begin{table}
  \caption{Systematic uncertainty for the absolute determination of the resonance strength in Sec.~\ref{sub:AbsoluteStrength}.\label{table:UncertaintiesAbsolute}}
  \begin{ruledtabular}
  \begin{tabular}{lD{.}{.}{-1}}
    \multicolumn{1}{c}{Uncertainty}				& \multicolumn{1}{c}{Contribution} \\
    \hline\noalign{\smallskip}
    Stoichiometry~\cite{Schmidt2013}				& 5.9\%	\\
    $\gamma$-ray angular distribution~\cite{Kozub1977}		& 3.8\%	\\
    $\gamma$-ray detection efficiency~\cite{Schmidt2013} 	& 2.3\%	\\
    Stopping power~\cite{Ziegler2010}				& 1.4\%	\\
    Beam current~\cite{Schmidt2013}				& 1.0\%	\\
    \hline\noalign{\smallskip}
    Total systematic uncertainty				& 7.6\%	\\
  \end{tabular}
  \end{ruledtabular}
\end{table}

\subsection{Ratio of the strengths of the 1.842\,MeV $\bm{^{40}}$Ca(p,$\bm{\gamma}$)$\bm{^{41}}$Sc and 4.5\,MeV $\bm{^{40}}$Ca($\bm{\alpha}$,$\bm{\gamma}$)$\bm{^{44}}$Ti resonances}
\label{sub:RelativeStrength}

The present absolute strength (Sec.~\ref{sub:AbsoluteStrength}) was determined on two targets which were also used to study the resonance triplet at $E_\alpha$ = 4.5\,MeV in the $^{40}$Ca($\alpha$,$\gamma$)$^{44}$Ti reaction. Therefore, as an alternative to the absolute resonance strength determination described in the previous section, the ratio of strengths of this resonance triplet and the resonance under study here has been determined. 

To calculate a ratio of resonance strengths, according to Eq.~(\ref{eq:Ratios_omega_gamma}) only the ratio of the two effective stopping powers has to be known, the absolute effective stopping power is not needed. The strength ratio depends only negligibly on the stoichiometric ratio, hence instead of 5.9\% for the stoichiometry and 1.4\% for the stopping power uncertainty, only 3.6\% for the ratio of stopping powers between proton beam and $\alpha$ beam~\cite{Ziegler2010} have to be included in the error budget. Likewise, the beam current normalization cancels out.

For the present ratio, in order to simplify the error calculation, only the ($\alpha$,$\gamma$) resonance strength determined by the activation method~\cite{Schmidt2013} is used. The ($\alpha$,$\gamma$) strength had also been determined previously using in-beam $\gamma$-ray spectrometry, but due to a poorly known angular distribution the results are less precise~\cite{Schmidt2013} and not used here.

From the ($\alpha$,$\gamma$) resonance strength, the following contributions to the error budget have to be taken into account: 1.1\% for the finite target thickness correction, 0.5\% for the $^{44}$Ti half-life~\cite{Ahmad2006}, and for the offline $\gamma$-ray counting 1.5\% for the detection efficiency and 2.5\% statistics~\cite{Schmidt2013}. From the (p,$\gamma$) resonance strength, contributions of 2.3\% for the efficiency in the in-beam $\gamma$-ray detection, 3.8\% for the 2882-keV $\gamma$-ray angular distribution~\cite{Kozub1977}, and 4.4\% for the result of the fit procedures (Table~\ref{table:ResonanceStrengths}) are included.

The total relative uncertainty for the resonance strength ratio is then 8\%, and the value obtained here is:
\[
  \frac{\omega\gamma({\rm p},\gamma)}{\omega\gamma(\alpha,\gamma)} = 0.0229\pm0.0018\nonumber
\]
Using $\omega\gamma(\alpha,\gamma)=8.4$\,eV~\cite{Schmidt2013}, an absolute strength of $\omega\gamma({\rm p},\gamma)=0.192$\,eV is obtained, confirming the result of the absolute determination of the resonance strength in Sec.~\ref{sub:AbsoluteStrength}.

\section{Discussion}
\label{sec:Discussion}

\begin{figure}
 \includegraphics[width=1.0\linewidth]{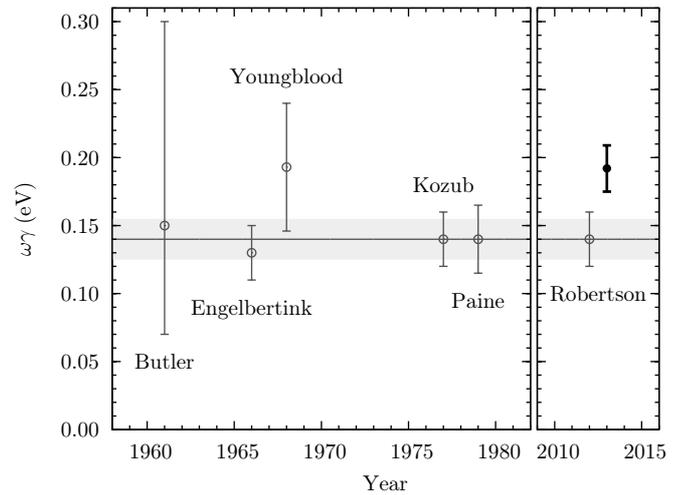}
 \caption{Literature data (open circles), compared with the present result (filled circle). The previous evaluated value~\cite{Endt90-NPA,Cameron2001} is given as a horizontal shaded bar.}
 \label{fig:Literature}
\end{figure}

The present new value of the resonance strength is significantly (3.5 times the uncertainty of the evaluated value) higher than the evaluated strength of $0.140\pm0.015$~\cite{Endt90-NPA,Cameron2001}. Only the values by Engelbertink and Endt~\cite{Engelbertink1966}, Kozub {\it et al.}~\cite{Kozub1977}, and Paine and Sargood~\cite{Paine1979} had been included in the evaluation.

However, the present new value is consistent within 1$\sigma$ error bars with the previous results by Butler~\cite{Butler1961} and Youngblood {\it et al.}~\cite{Youngblood1968}, and consistent within 2$\sigma$ with all the other works when taken individually~\cite{Engelbertink1966,Kozub1977,Paine1979,Robertson2012} (Fig.~\ref{fig:Literature}). 

In order to understand the discrepancies, it is instructive to consider the target materials used. It has to be noted that metallic calcium targets may easily oxidize and transform to CaO or even Ca(OH)$_2$. In case of an incorrectly determined stoichiometry for calcium, one would therefore expect an underestimate of the effective stopping power and, thus, an underestimate of $\omega\gamma$ for a given experimentally determined yield. Such an effect may well explain why several of the works using metallic calcium or calcium oxide~\cite{Kozub1977,Paine1979} give lower strength values than the present one.

The early work of Engelbertink and Endt~\cite{Engelbertink1966} was part of a network of intercomparison of various resonance strengths, including the $E_{\rm p}$ = 588\,keV resonance in the $^{32}$S(p,$\gamma$)$^{33}$Cl reaction. The strength of this resonance has later been re-studied and seems to be a factor of 2 higher~\cite{Iliadis92-NPA}, which would bring the Engelbertink and Endt~\cite{Engelbertink1966} result in agreement with the present value.

Two early works devoted particular attention to the target composition. Butler~\cite{Butler1961} electrodeposited calcium on the backing and then deliberately oxidized it to form calcium oxide. Subsequently, the targets where ignited to white heat with a torch, removing impurities and ensuring that the target material remains burnt lime (CaO). Youngblood {\it et al.}~\cite{Youngblood1968} used metallic calcium targets, and a detailed description of the procedures used is available~\cite{Youngblood1965}. Calcium metal was scraped, removing surface oxidation, before the target was evaporated in situ. A proton elastic scattering experiment resulted in an oxygen content of less than 2\%. The $\omega\gamma({\rm p},\gamma)$ values of Butler~\cite{Butler1961} and Youngblood {\it et al.}~\cite{Youngblood1968} are in agreement with the present one.

Much less details on target preparation and handling are available in the works by Kozub {\it et al.} \cite{Kozub1977}, Paine and Sargood \cite{Paine1979}, and Robertson {\it et al.} \cite{Robertson2012}. Kozub {\it et al.} \cite{Kozub1977} state that they used the $^{16}$O(p,$\gamma$)$^{17}$F reaction to determine the oxygen content of their targets, and that corrections to the effective stopping amounted to a few percent. However, no details on which of the three $^{16}$O(p,$\gamma$)$^{17}$F transitions were actually used for the analysis and on the angular corrections (which are significant in this case) are given, hence it is possible that Kozub {\it et al.} underestimated the oxygen content of the target~\cite{Kozub1977}. For the other two works by Paine and Sargood \cite{Paine1979} and by Robertson {\it et al.} \cite{Robertson2012}, even less details are available, so target oxidation cannot be excluded there, either. 

The present target composition has been determined both by the elastic recoil detection (ERD) method and by nuclear reactions~\cite{Schmidt2013}. It is consistent with fully oxydized and hydrogenated calcium.

The correctness of the present new result is further corroborated by the relative resonance strength determination, which to very good approximation is independent of stoichiometry. The strength of the 4.5\,MeV resonance triplet in the $^{40}$Ca($\alpha$,$\gamma$)$^{44}$Ti reaction is now rather well-determined. The normalization value used here, $\omega\gamma(\alpha,\gamma)=8.4\pm0.6$\,eV~\cite{Schmidt2013} is the most precise value available, but other previous resonance strengths~\cite{Dixon1980,Nassar2006,Vockenhuber2007} that have been determined without reference to $^{40}$Ca(p,$\gamma$)$^{41}$Sc are all close to it. An exception is the work by Robertson {\it et al.}, where the $^{40}$Ca($\alpha$,$\gamma$)$^{44}$Ti resonance strength at 4.5\,MeV has been determined relative to the $^{40}$Ca(p,$\gamma$)$^{41}$Sc resonance strength under study here \cite{Robertson2012}.  

\section{Summary}

The strength of the $E_{\rm p}$ = 1842\,keV resonance in the $^{40}$Ca(p,$\gamma$)$^{41}$Sc reaction has been re-measured. The result, $\omega\gamma=(0.192\pm0.017)$\,eV, is higher than the value from a previous evaluation \cite{Endt90-NPA,Cameron2001}. In addition, the ratio of strengths of the latter resonance and the 4.5\,MeV resonance triplet in the $^{40}$Ca($\alpha$,$\gamma$)$^{44}$Ti reaction has been determined to be $0.0229\pm0.0018$. The present value may be used in the future as a normalization point in coming precision experiments on $^{40}$Ca targets.

\begin{acknowledgments}
The excellent support by the staff and operators of the HZDR ion beam center, by Andreas Hartmann (HZDR), and by Bettina Lommel and the GSI target laboratory is gratefully acknowledged. --- This work was supported in part by DFG (BE4100/2-1), by HGF (NAVI, VH-VI-417), by EuroGENESIS, and by the European Union (FP7-SPIRIT, contract no. 227012).
\end{acknowledgments}

\end{document}